\documentclass[pdflatex,sn-nature]{sn-jnl}

\usepackage{graphicx}%
\usepackage{multirow}%
\usepackage{amsmath,amssymb,amsfonts}%
\usepackage{amsthm}%
\usepackage{mathrsfs}%
\usepackage[title]{appendix}%
\usepackage{xcolor}%
\usepackage{textcomp}%
\usepackage{manyfoot}%
\usepackage{booktabs}%
\usepackage{algorithm}%
\usepackage{algorithmicx}%
\usepackage{algpseudocode}%
\usepackage{listings}%
\hypersetup{hypertexnames=false}

\usepackage[colorinlistoftodos,prependcaption, textwidth=2cm,,textsize=scriptsize]{todonotes}

\theoremstyle{thmstyletwo}%

\theoremstyle{thmstylethree}%

\title{Whole Slide Concepts: A Supervised Foundation Model For Pathological Images}

\begin{document}


\author*[1]{\fnm{Till} \sur{Nicke}}\email{till.nicke@mevis.fraunhofer.de}
\author[1]{\fnm{Daniela} \sur{Schacherer}}
\author[1]{\fnm{Jan Raphael} \sur{Schäfer}}
\author[2,4]{\fnm{Natalia} \sur{Artysh}}
\author[2,4]{\fnm{Antje} \sur{Prasse}}
\author[1]{\fnm{André} \sur{Homeyer}}
\author[1,3]{\fnm{Andrea} \sur{Schenk}}

\author[1]{\fnm{Henning} \sur{Höfener}}
\equalcont{These authors contributed equally to this work.}

\author[1]{\fnm{Johannes} \sur{Lotz}}
\equalcont{These authors contributed equally to this work.}

\affil[1]{\orgdiv{Institute for Digital Medicine MEVIS}, \orgname{Fraunhofer}, \orgaddress{\city{Bremen/Lübeck/Hannover}, \country{Germany}}}

\affil[2]{\orgdiv{Institute for Toxicology and Experimental Medicine}, \orgname{Fraunhofer}, \orgaddress{\city{Hannover}, \country{Germany}}}
\affil[3]{\orgdiv{Institute for Diagnostic and Interventional Radiology}, \orgname{Hannover Medical School}, \orgaddress{\city{Hannover}, \country{Germany}}}
\affil[4]{\orgdiv{Department of Pulmonology}, \orgname{Hannover Medical School}, \orgaddress{\city{Hannover}, \country{Germany}}}



\abstract{Foundation models (FMs) are transforming computational pathology by offering new ways to analyze histopathology images. However, FMs typically require weeks of training on large databases, making their creation a resource-intensive process. 
In this paper, we present a training for foundation models from whole slide images using supervised, end-to-end, multitask learning on slide-level labels. Notably, it is the first model to incorporate cancer subtyping, risk estimation, and genetic mutation prediction into one model. The presented model outperforms self-supervised models on seven benchmark tasks while the training only required 5\% of the computational resources. The results not only show that supervised training can outperform self-supervision with less data, but also offer a solution to annotation problems, as patient-based labels are widely available through routine clinical processes.
Furthermore, an attention module provides a layer of explainability across different tasks and serves as a tumor detector for unseen cancer types. To address the issue of closed-source datasets, the model was fully trained on openly available data. The code and model weights are made available under \url{https://github.com/FraunhoferMEVIS/MedicalMultitaskModeling}.}

\keywords{end-to-end learning; multiple-instance learning; foundation model; weakly-supervised learning}



\maketitle

\section{Introduction}\label{sec1}

Traditionally, diagnosing whole-slide images (WSIs) is often limited by the volume of cases, which are predicted to grow in the future \cite{weir2021cancer, sung2021global}. Foundation Models (FMs) have the potential to transform the field of computational pathology by providing a new way to analyze digital histopathology images and support the decreasing workforce of pathologists. The promise to require only few training images to generalize well across various tasks and centers offers the possibility of focused research in data sparse domains, and to diagnose slides in an automated and fast manner. 

The training of FM is mainly driven by self-supervised learning (SSL), which provides a label-free training method for extracting robust and generalizable features from images \cite{gui2024survey}. To improve the learned representations, various SSL methods were developed from contrastive learning, in which different views of the same image are represented as similar latent vectors, to masked image modeling, in which smaller patches of the input image are obscured and are to be reconstructed by the model \cite{gui2024survey}. For WSI-based FM, a two-step training process is widely adopted. First, a pre-trained, frozen patch encoder extracts patch-level features. Then, in the second step, these features are further processed to encode slide-level embeddings \cite{ding2024multimodal,lenz2024unsupervised,provgigapath,shaikovski2024prism}.

Recent advances apply this scheme, such as Xu et al. \cite{provgigapath}, who suggest to use masked image modeling in latent space to learn a transformer based patch aggregator. Other approaches include deploying multiple frozen patch encoders to generate different views of the same patch and using a contrastive learning paradigm as an aggregation strategy, presented by Lenz et al. \cite{lenz2024unsupervised}. Wang et al. \cite{wang2024pathology} also use a frozen patch encoder in combination with different organ origins of slides in a CLIP learning paradigm.

However, these approaches rely on frozen patch patch encoders to solve WSI representations using a patch-encoder method that is pre-trained using SSL. This has consequences on training time and research cycles, as SSL-based methods tend to converge slower, compared to supervised methods. Thus, self-supervised training of two or more separate model components requires substantial computing power, a large amount of data, and does not necessarily yield optimal models \cite{tang2025revisiting}. On the other hand, recent research suggests that learning from slide-level labels in an end-to-end manner is possible and can even outperform two-step methods \cite{jiang2023mhattnsurv,pinckaers2021detection,tang2025revisiting}. 

To date, end-to-end training approaches have demonstrated strong performance on single, individual tasks \cite{tang2025revisiting, jiang2023mhattnsurv, pinckaers2021detection}. Nonetheless, multi-task learning has been shown to be an efficient strategy for training foundation models, requiring only a fraction of data and computational resources \cite{schafer2024overcoming, nicke2025tissue}. Using available endpoints, such as patch labels or segmentations, an FM can be trained to simultaneously solve various tasks in digital pathology and remain adaptable to new problems \cite{nicke2025tissue}. Incorporating slide-level labels in an end-to-end manner into the multi-task learning pipeline was previously unexplored, as large labeled data-cohorts were unavailable for supervised training. This lack of labels can be overcome by taking advantage of slide- or patient-level labels that are broadly available in routine diagnostics.

In this paper we ,therefore, investigate the performance of models trained using weakly labeled, end-to-end, multi-task learning and compare them to SSL-based counterparts. Through joint optimization of various tasks, a model is optimized to solve different clinically relevant problems from weakly labeled WSIs. This approach alleviates the need for specific annotations in supervised training, while at the same time drastically reducing the compute resources needed to train a WSI-based FM.

We draw on a large data cohort with readily available endpoints for WSIs that can be accessed through the National Cancer Institute (NCI) Imaging Data Commons (IDC). This open-source, cloud-based data repository contains cancer data from various modalities, including brightfield and fluorescence slide microscopy, as well as the corresponding endpoints for the given WSIs \cite{fedorov2023national,schacherer2023nci,bontempi2024end}. Building up on open-source data facilitates reproducibility and bias exploration of published models. 
As the training scheme captures general concepts from various slides, we term it Whole Slide Concepts (WSC). Our contributions can be summarized as follows: 
\begin{itemize}
\item WSC exhibits state-of-the-art performance in various tasks while relying on fewer data and a smaller model architecture compared to other SSL-based FMs.
\item We show that end-to-end multitask learning on slide-level labels is resource-efficient and enables the joint optimization of the patch and slide encoder.
\item Through exclusively training with openly available data, primarily from the NCI IDC \cite{fedorov2021nci}, the training is fully reproducible.
\end{itemize}


\section{Results}\label{sec2}
We measure the training time, in GPU hours, and show that WSC, with a Swin-Transformer V2 tiny backbone, uses significantly fewer resources during training. Table \ref{tab:run_time} compares the reported hours of training and estimated kWh used of WSC, with a Swin-tiny backbone \cite{liu2022swin} (WSC-tiny), Prov-GigaPath \cite{shaikovski2024prism}, and CHIEF \cite{wang2024pathology} using the max wattage of reported GPUs. We further evaluated the WSC model on 7 different benchmark tasks, listed in Table \ref{tab:eval_tasks}, and compared it to the WSI-based foundation models CHIEF \cite{wang2024pathology}, Prov-GigaPath \cite{provgigapath}, and the patch-level encoder UNI \cite{chen2024towards}. 

The models' performances are tested on four different cancer subtype classification tasks, which are equally divided into in- and out-of-domain tasks based on their similarity to the pretraining data distribution (Table \ref{tab:eval_tasks}). As metric, the average balanced accuracy and area under the receiver operating characteristic curve (AUC) over four distinct test runs for the corresponding tasks is reported. 
Additionally, we test the performance on three risk estimation tasks for overall survival on unseen benchmark datasets, using the average c-index over four runs as an evaluation metric.
Further, we analyze WSC in terms of explainability on unseen data. The predicted attention scores of the model are used to determine cancerous regions within unseen WSIs and compared to expert annotations.  

To address questions regarding the influence of patch- and slide-based labels, we compare two versions of our model. Both models comprise a Swin-tiny \cite{liu2022swin} backbone, one of which is trained using slide-level labels only, while the other is trained using both, patch- and slide-based labels. To account for parameter scaling experiments, a third model is trained with a Swin-small \cite{liu2022swin} backbone using patch- and slide-level labels. In the results, these models are mentioned as WSC-tiny$_{wsi}$, WSC-tiny$_{wsi+patch}$, and WSC-small$_{wsi+patch}$, respectively.


\subsection{WSC requires fewer resources}
\begin{table}[tbp]
    \caption{Comparison of GPU hours and estimated kWh used during training of different foundation models.}
    \label{tab:run_time}
    \centering
    \begin{tabular}{l|c|c|c|c|c}
        \toprule
        Model & GPU & Max Watt & Num GPUs & Total GPU H. & kWh\\
        \midrule
    Prov-GigaPath \cite{provgigapath} enc. & A100 & 400 & - & - & - \\
    Prov-GigaPath \cite{provgigapath} aggr. & A100 & 400 & 4 & 3072 & 1229 \\
    \midrule
    CTransPath \cite{wang2022transformer} & V100 & 300 & 48 & 12000 &  3600\\
    CHIEF \cite{wang2024pathology} & V100 & 300 & 8 & - & - \\
    \midrule
    Tissue Concepts \cite{nicke2025tissue} & RTX A5000 & 230 & 1 & 160 & 37 \\
    WSC-tiny & A100 & 400 & 1 & 500 & 200\\
    \botrule
    \end{tabular}
\end{table}
Supervised, MTL-based training uses substantially fewer resources compared to self-supervised training, which we regard as especially important with a focus on the environmental impact of current developments of larger foundation models.

Table \ref{tab:run_time} shows a comparison of different run times, taken from the reported literature of other WSI-based foundation models \cite{provgigapath, wang2022transformer, wang2024pathology}. In \cite{wang2022transformer}, the authors report training the CTransPath patch-encoder for 250 hours on 48 V100 GPUs, which results in a total of 12000 GPU hours and an energy usage of approximately 3600 kWh. Being based on CTransPath, the training of CHIEF is not described in further detail concerning energy footprint in \cite{wang2024pathology}, which is why there is no complete estimation for CHIEF. In \cite{provgigapath}, the authors report the training time and configuration for the aggregation module, which is a total of 3072 A100 hours. This is the equivalent of around 1230 kWh of energy used. However, the authors do not report the pretraining of the patch encoder for Prov-GigaPath. Therefore, our estimate is only a lower bound of the actual energy consumption.

We trained the WSC model for 500 hours on one single A100, with a maximum power consumption of 400W. Under the assumption of the GPU being under full load all the time, this results in an energy usage of 200 kWh for the training. Our previous patch-level encoder Tissue Concepts \cite{nicke2025tissue} was trained using 37 kWh and its weights served as the starting point for our model training. Overall, this results in an energy usage of 237 kWh for the whole foundation model training, which corresponds to 16\% (Prov-GigaPath) or 5\% (CHIEF) of the energy used in SSL-based model training.

\subsection{WSC shows state of the art performance for in-domain tasks}

\begin{figure}
    \centering
    \includegraphics[width=0.99\linewidth]{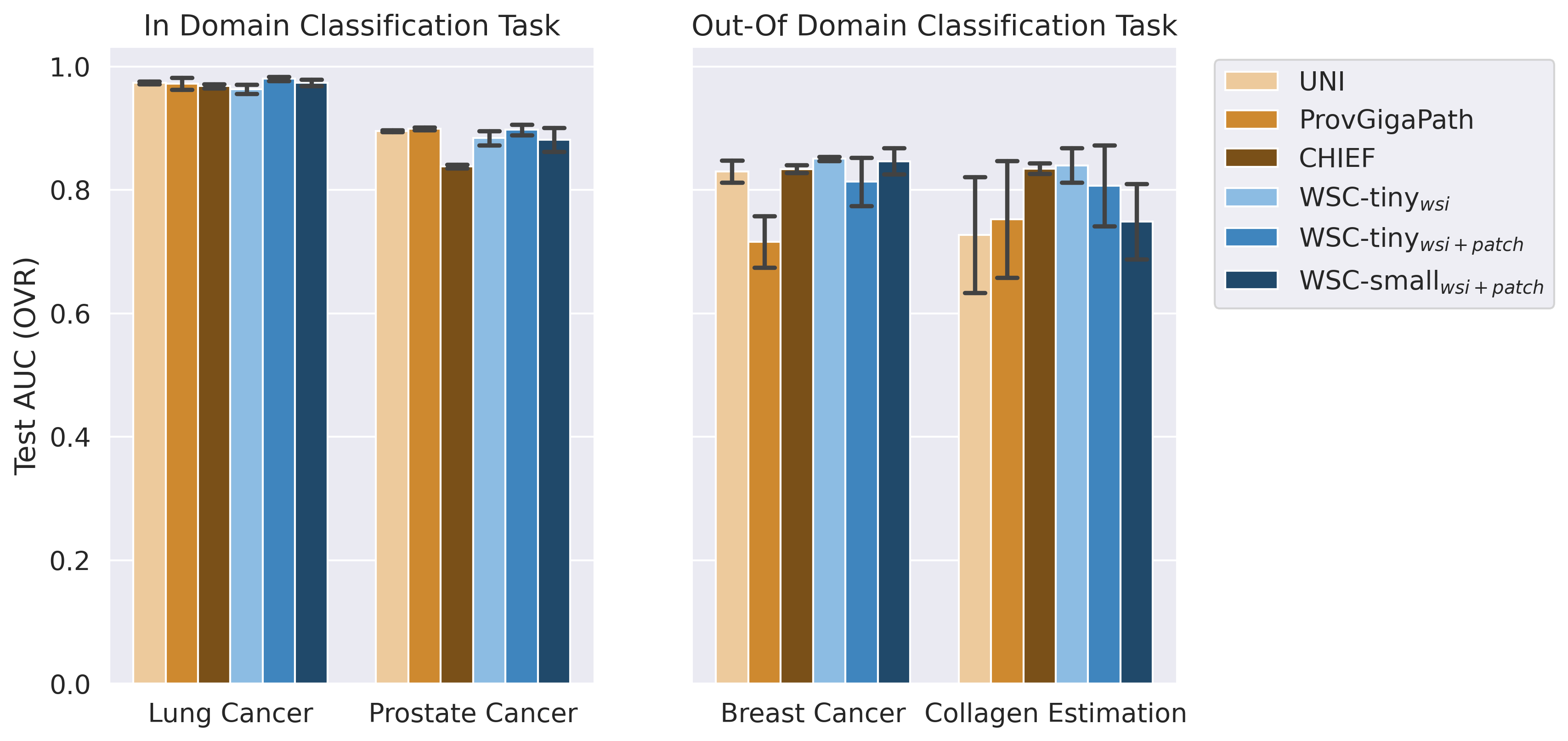}
    \caption{Average AUC over four runs on four cancer subtyping tasks divided into in- (left) and out-of-domain (right) tasks depending on their similarity to the training distribution.}
    \label{fig:ClfResults}
\end{figure}

To evaluate the models' performance on in-domain tasks, we computed the mean performance over four runs on the non-small cell lung cancer (NSCLC) subtyping task (CPTAC LUAD vs CPTAC LUSC) and on the ISUP grading task in prostate cancer from the PANDA dataset (see \ref{sec:eval_prot}). 
For NSCLC, models were trained on the respective datasets from TCGA dataset and evaluated on data from CPTAC, providing a cross-cohort evaluation (Figure \ref{fig:ClfResults} (Lung Cancer)).

When subtyping NSCLC, WSC-tiny$_{wsi+patch}$ outperforms CHIEF \cite{wang2024pathology}, despite both models sharing the same backbone architecture (AUC: 0.98 vs 0.97, p $<$ 0.05). WSC-tiny$_{wsi+patch}$ also outperforms Prov-GigaPath \cite{provgigapath} on this task (AUC: 0.98 vs 0.97), even though Prov-GigaPath is reported to have a much larger parameter count (ca. 28M versus ca. 1.1B \cite{campanella2025clinical, provgigapath}). Between the different model variants, WSC-tiny$_{wsi+patch}$, the combined patch- and slide-level label model, outperforms the model solely trained on WSI-level labels (AUC: 0.98 vs 0.96, p WSC$<$ 0.01). Scaling up the parameter count did not lead to a performance increase on this task, as  WSC-small$_{wsi+patch}$ did not achieve higher scores compared to WSC-tiny$_{wsi+patch}$ (AUC: 0.98 vs 0.97).

ISUP grading performance was evaluated on the PANDA challenge dataset, which was held out from training and used as an external test set. We nevertheless consider this task in-domain, as WSC was pre-trained to predict ISUP grading on TCGA-PRAD.

In ISUP prediction, WSC-tiny$_{wsi+patch}$ performs on par with UNI and Prov-GigaPath with an AUC of 0.89 and a balanced accuracy of 0.57. WSC-tiny$_{wsi+patch}$ performed superior to CHIEF, which uses the same architectural backbone (AUC: 0.89 vs 0.83, p $<$ 0.001).
The patch- and WSI-based model also outperforms its WSI-only counterpart (AUC: 0.89 vs 0.84). No increase in performance can be seen by increasing the parameter count (AUC: Tiny: 0.89 vs Small: 0.88) (Figure \ref{fig:ClfResults} (Prostate Cancer)).

\subsection{WSC learns transferrable features for out-of-domain tasks}

To evaluate the models' out-of-domain performance, two unseen tasks, Breast Cancer subtyping and Lung Fibrosis classification, were selected (Figure \ref{fig:ClfResults}, right).

Breast Cancer corresponds to the BReAst Cancer Subtyping (BRACS) dataset. On BRACS, WSC-tiny$_{wsi}$ achieves the best mean AUC (0.85), outperforming CHIEF and UNI (AUC: 0.83 each) as well as Prov-GigaPath (AUC: 0.71; p $<$ 0.01 versus all other models).
Adding patch-level supervision slightly decreases the performance on this task: WSC-tiny$_{wsi+patch}$ reaches an AUC of 0.81. Increasing the backbone size does not improve performance, as WSC-small$_{wsi+patch}$ attains an AUC of 0.84.

The lung fibrosis task is based on an internal dataset comprising 94 H\&E-stained precision-cut lung slices (PCLS) with corresponding collagen quantification (see Section \ref{subsec:eval} for details). The goal is to classify the cases' amount of fibrosis into high- versus low-fibrotic. WSC-tiny$_{wsi}$ achieves the best AUC (0.84), followed by CHIEF (0.83). Prov-GigaPath, UNI, and WSC-small$_{wsi+patch}$ perform worse with AUC of 0.75, 0.72, and 0.74, respectively. Adding patch-level supervision decreases performance (WSC-tiny$_{wsi}$: 0.84 vs WSC-tiny$_{wsi+patch}$: 0.80).


\subsection{WSC estimates risk from slides}
\begin{figure}
    \centering
    \includegraphics[width=0.95\linewidth]{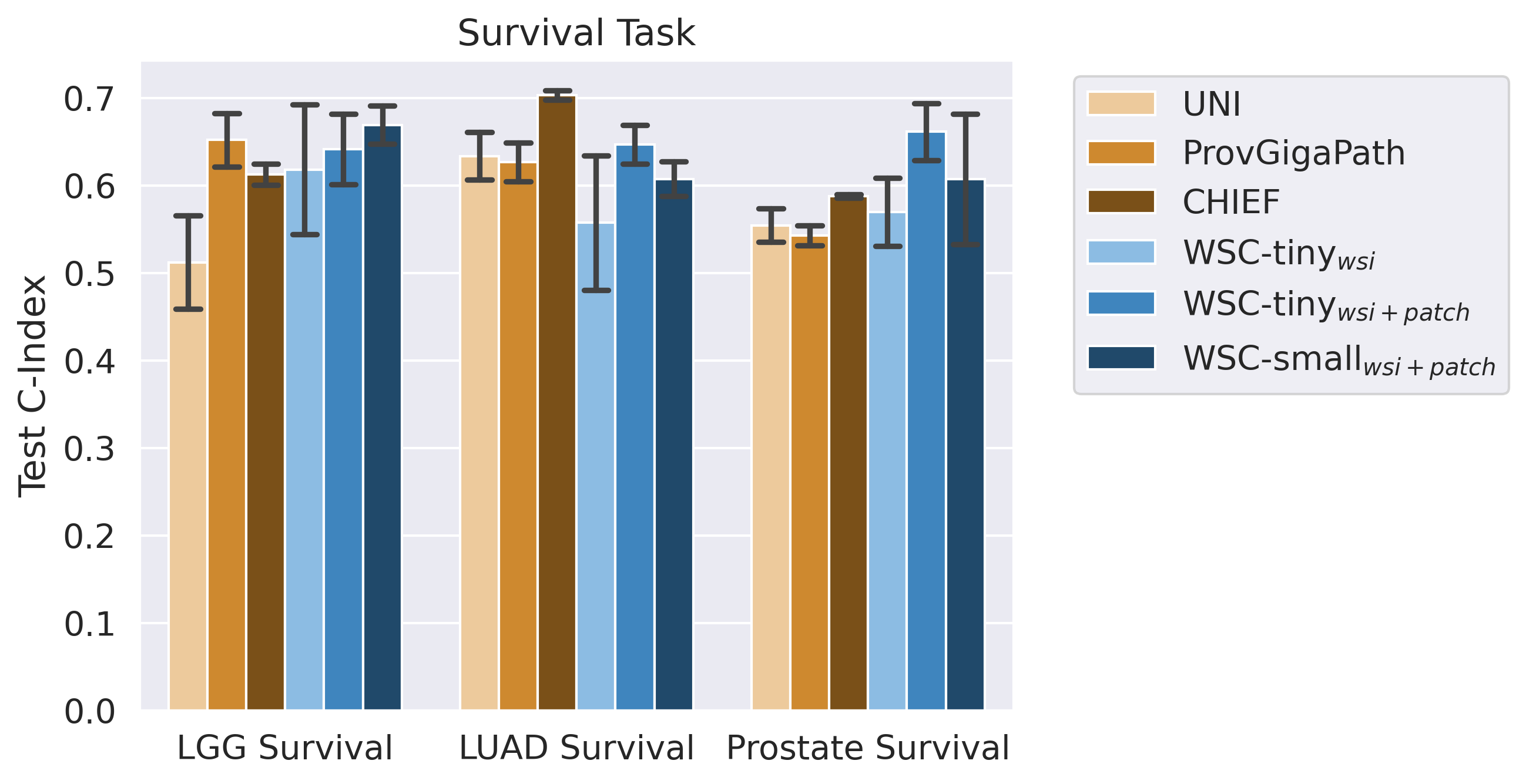}
    \caption{Mean c-index of four distinct runs on three benchmark evaluation tasks to estimate OS on Brain, Lung, and Prostate tissue (left to right).}
    \label{fig:survivalResults}
\end{figure}
We evaluate the models' performances on three risk estimation tasks. The averaged c-index is presented in Figure \ref{fig:survivalResults}. LGG Survival is based on the low-grade glioma data cohort from the TCGA. LUAD Survival uses the LUAD cohort from CPTAC as unseen LUAD cases for risk estimation. Prostate Survival is based on the dataset provided by the Leopard challenge.

The results for the risk estimation of low-grade glioma show that WSC-tiny$_{wsi+patch}$ outperforms the WSI-only model WSC-tiny$_{wsi}$ (c-index: 0.64 vs 0.61). WSC-tiny$_{wsi+patch}$ also outperforms CHIEF, which offers the same parameter count (0.64 vs 0.61). The patch-based encoder UNI was not able to solve this task, as presented by the c-index of 0.51. Scaling up the parameter count in the MTL scenario led to performance improvements. WSC-small$_{wsi+patch}$ showed performances superior to the smaller backbone (c-index: 0.66 vs. 0.64). It also significantly outperformed UNI and CHIEF (p $<$ 0.001).

When estimating risk for CPTAC-LUAD, all models show major performance differences. CHIEF performed best with a c-index of 0.70. WSC-tiny$_{wsi+patch}$ performed second best with an average c-index of 0.64. WSC-tiny$_{wsi}$ was outperformed by the other models with the lowest c-index of 0.55 in this task. WSC-small$_{wsi+patch}$ was outperformed by Prov-GigaPath and UNI (c-index: 0.60 vs 0.62 vs 0.63, respectively). Increasing the parameter count did not lead to an increase in performance.

Estimating the risk for prostate cancer was selected as a third survival task. For this, the dataset from the Leopard challenge was used. WSC-tiny$_{wsi+patch}$ outperformed the other FM with a c-index of 0.66 (p $<$ 0.01 for all). Chief, Uni, and Prov-GigaPath scored 0.58, 0.54, and 0.55, respectively. WSC-tiny$_{wsi+patch}$ also outperformed WSC-tiny$_{wsi}$ and WSC-small$_{wsi+patch}$ with c-index of 0.56 and 0.60, respectively. As seen before, the increase in parameter count did not lead to a better performance.

\subsection{WSC is a general tumor detector}
Next to quantitative performances on benchmarks, we found that WSC-tiny$_{wsi+patch}$ can be used as a general tumor detector on unseen data. For the evaluation, we used the TUPAC 16 challenge dataset, as it contains annotated tumor regions and was not used during the training of the models \cite{veta2019predicting}. 

\begin{figure}
    \centering
    \includegraphics[width=0.49\linewidth]{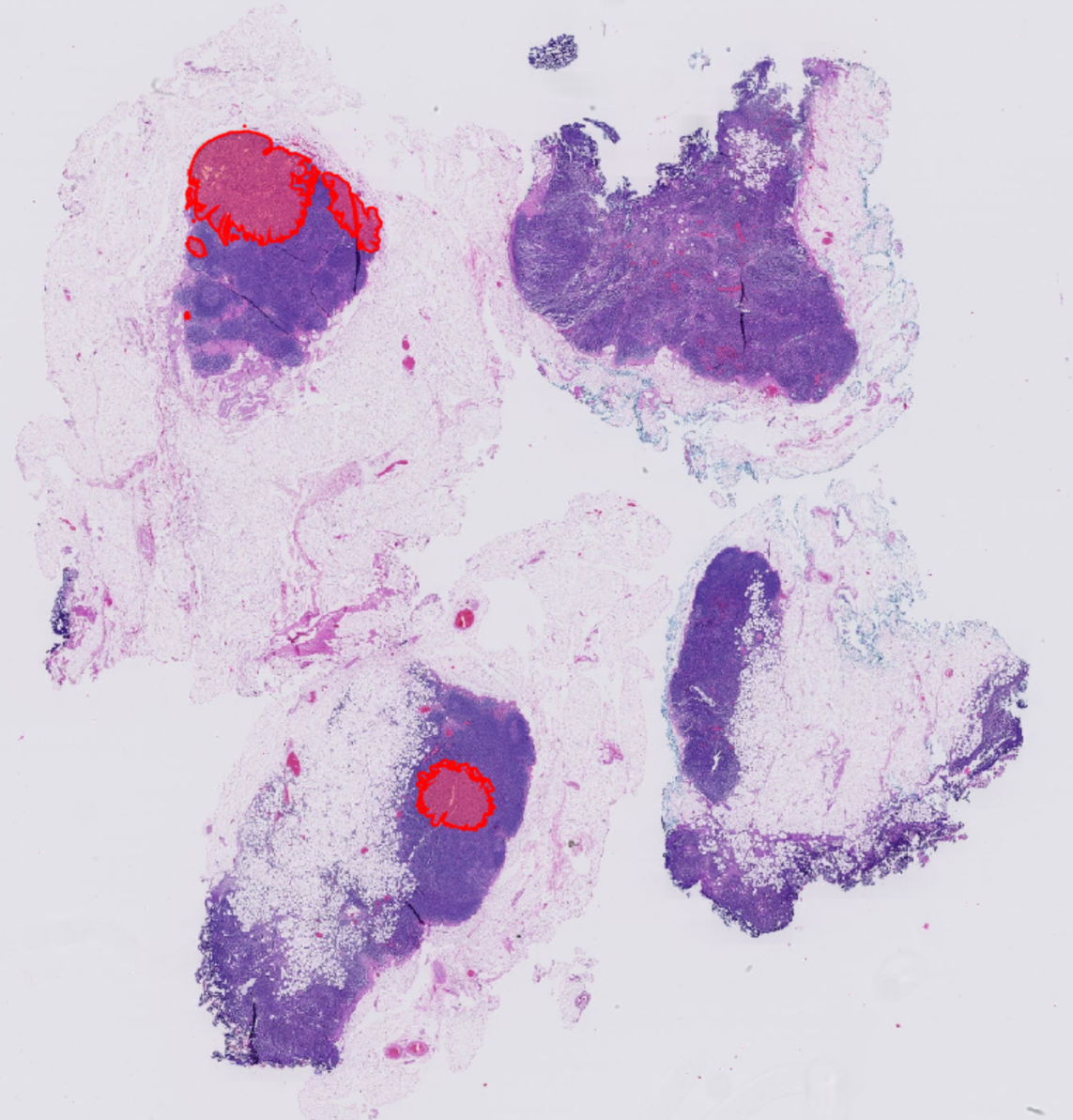}
    \includegraphics[width=0.49\linewidth]{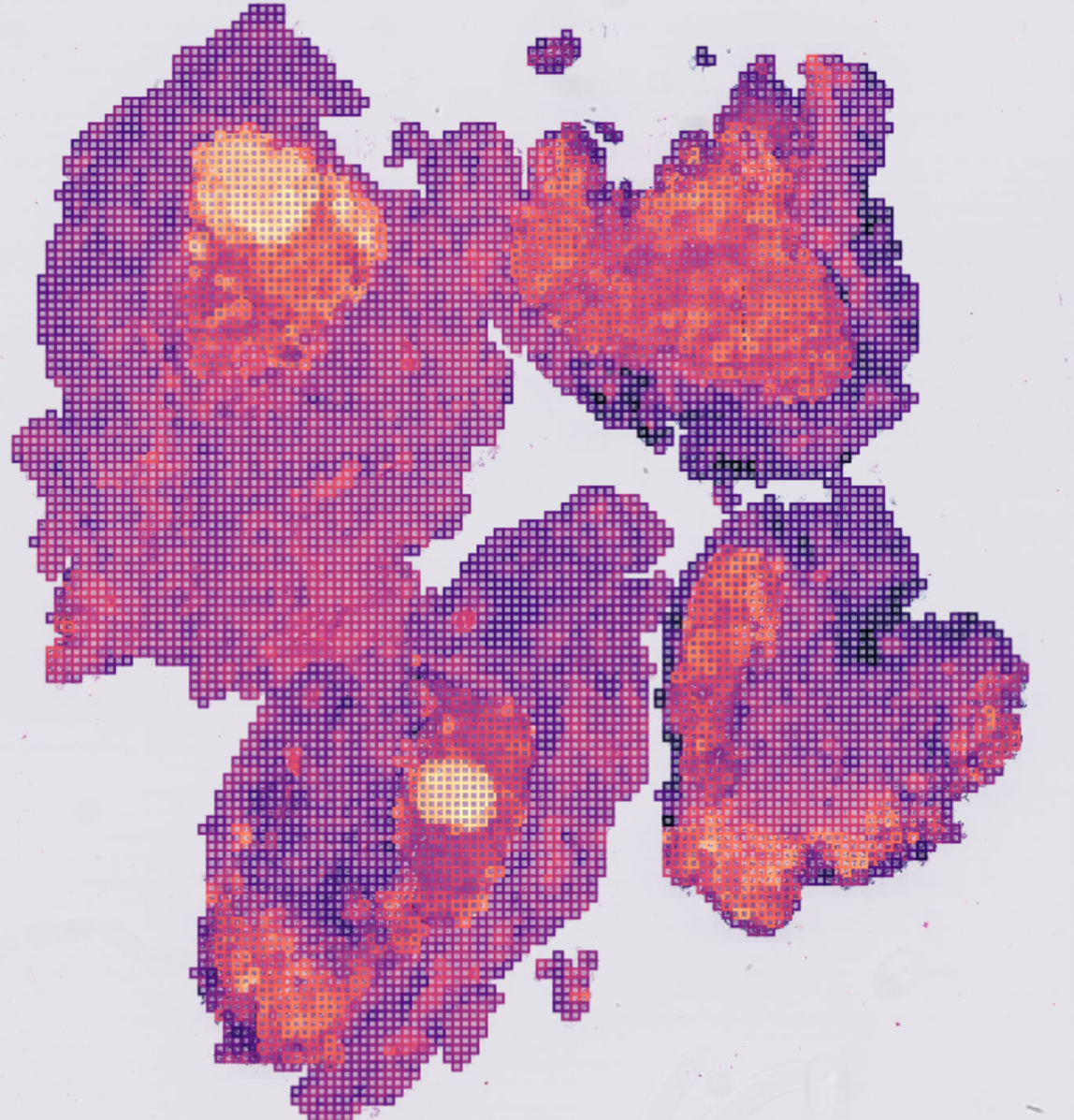}
    \caption{Side-by-side comparison of the tiles with the highest attention values (right) and the expert tumor annotations (left) without fine-tuning of WSC on an unseen slide from the TUPAC16 cohort.}
    \label{fig:tumor_detection_tupac}
\end{figure}

Through the multi-head attention (MHA) pooling module (Figure \ref{fig:figure_1}), the relative patch importance that is estimated by the model for each WSI can be computed, providing a layer of explainability.
We compare the WSC-tiny$_{wsi+patch}$ attention values with the expert tumor annotations from the TUPAC16 challenge dataset. Figure \ref{fig:tumor_detection_tupac} shows an example of the expert annotations (left) and the overlayed attention weights (right). We find that the tiles that are within the annotated tumor regions receive higher attention scores from the model compared to background or non-annotated foreground regions. 

For non-tumor-related tasks, such as fibrosis estimation, a similar pattern can be observed. As can be seen in Figure \ref{fig:attention_weights}, tiles obtained higher attention values by the model if they showed signs of collagen, such as wave-like structures. It is to be noted that the model was fine-tuned to estimate the collagen area fraction and visualized the attention weights in Figure \ref{fig:attention_weights}. The network effectively highlights areas of high collagen in H\&E-stained images of PCLS.


\section{Discussion}\label{sec12}
In this paper, we show that end-to-end learning using weakly supervised labels is an efficient approach for creating whole slide image foundation models. Even though fewer computing resources were used, the models exhibit similar performance compared to models with more parameters or those that were trained with more data. By solely using openly available data, the reproducibility and bias research for the WSC models is enhanced compared to models that rely on closed-source data.

The training of WSC demonstrates that self-supervised training, while creating robust and capable models, is less efficient for creating vision foundation models in digital pathology. Comparing the WSC models against other WSI-based foundation models showed that WSC outperforms the other models in almost all of the benchmark tasks while relying on fewer compute resources and available data. In \cite{provgigapath}, the authors mention that their curated dataset comprises more cases compared to the TCGA. Yet, the supervised pretraining strategy, presented in this paper, still outperforms the foundation model Prov-GigaPath while relying on fewer computational resources and data. This shows that scaling up data sets with resource-hungry model training may lead to underperforming models in the end. Using the expert knowledge present in public datasets not only yields better results but also requires fewer computing resources and less time, extending the research presented in \cite{nicke2025tissue}. To fully assess the capabilities of the supervised FM, a wider range of evaluation tasks need to be considered. However, since most benchmarks rely on openly available data, these tasks need to be carefully selected to avoid data leakage from the supervised training.

For a parameter-matched comparison, CHIEF was included as a baseline. Overall similar performances can be seen across in- and out-of domain cancer subtyping tasks. For more complex tasks, such as outcome prediction, WSC shows superior performance in two out of three tasks. However, high standard deviations on these tasks also signal potential instability for further applications. These could be the result of a chosen tile sampling strategy, where 300 patches per slide were extracted to fine-tune the respective model. Especially for more complex tasks, these patches do not need to be representative of the outcome. Additionally, the parameters of the TC models were optimized during training to estimate outcome prediction next to other tasks. Higher instability in transfer performance suggests suboptimal features of the models to accommodate more complex relationships. Improved training strategies should be investigated in the future.

We did not compare our model to other models that relied on multi-modal data, like TITAN \cite{ding2025multimodal}. Since no pathology reports or other forms of text data are available through the IDC or other public data sources, incorporating text as a domain into the training was not feasible to hold up to the open source requirements for the training. Training a uni-modal model, the best comparison was with other uni-modal models. However, text aligning the WSI representation of WSC with pathological reports, to obtain more expressive representations and extract report information form slides alone should be investigated in the future. A clear problem for this is the lack of available text-slide datasets and challenges for joint optimization of text and image encoder in an end-to-end manner.

When comparing the needed computational resources for training the individual models, TC shows greater efficiency in kWh needed. However, Table \ref{tab:run_time} shows some blanks, which we were not able to fill. These mainly stem from missing reports in training time or the specific GPU used during training. Therefore, the estimates for the models shown in the table might be too optimistic. However, we are confident that the trend demonstrated, namely that SSL-based methods require significantly higher computing power, is nevertheless of a general nature.  A different aspect of environmental impact lies in inference and potential retraining cycles. For inference, lower parameter counts are to be preferred, since they can be run on smaller and less power-hungry GPUs \cite{campanella2025clinical}. A different aspect lies in the potential retraining cycles, as new data becomes available. Here, task-specific fine-tuning of the chosen MTL approach can potentially lead to catastrophic forgetting of other tasks if they are not used during training \cite{bayasi2025beyond}. The exploration of continual learning \cite{kumari2025continual, bayasi2025beyond} with respect to task-specific fine-tuning while maintaining the generalization capabilities of the foundation model should be explored in this regard.

Reproducibility is facilitated by using openly available data and relying on fewer compute resources, such that the training of the model can be done without access to latest state of the art GPUs.
Other researchers can more easily investigate and counteract known or unknown biases inherent in the training data, such as the TCGA data mainly stems from North America and is therefore completely representative of the global population \cite{kheiri2025investigation}. Community-driven improvement in this regard is highly welcomed and encouraged. We, therefore, rely on versioning of used files in the IDC, which allows researchers to address issues like misclassifications of single files. For future reproducibility studies, the same files can be accessed. In the case of the PLCO prostate data, a short application is required. 

In general, future research should further focus on compressing foundation models to make them accessible in environments, that do not have access to larger GPU clusters, such as phones or standard work stations \cite{shi2025foundation}. By additionally, increasing the number and diversity of tasks, the MTL pipeline can solve, the model can be applied in various scenarious from biomarker discovery to clinical applications.

\section{Methods}\label{sec11}

\subsection{Training data}
\begin{table}[h!]
    \caption{Overview of different tasks created from the TCGA and PLCO cohorts.}
    \centering
    \begin{tabular}{c|c|c|c|c}
    Cohort & Target & Num WSIs & Organ & Origin\\
    \hline
    \hline
    Organ Subtype & Subtype & 22225 & Various & TCGA \\
    \hline
    SKCM & Overall Survival & 844 & Skin & TCGA\\
    \hline
    PLCO & Overall Survival & 2380 & Prostate & PLCO\\
    \hline
    LGG/GBM & Subtype & 3529 & Brain & TCGA\\
    \hline
    KICH/KIRP/KIRC & Subtype & 3207 & Kidney & TCGA\\
    \hline
    \multirow{2}*{PRAD} & Overall Survival & 525 & Prostate & TCGA\\
    & Subtype & 525 & Prostate & TCGA\\
    \hline
    \multirow{5}*{BRCA} & Overall Survival & 1023 & Breast & TCGA\\
    & Subtype & 1023 & Breast & TCGA\\
    & TP53 Mutation Prediction & 1023 & Breast & TCGA\\
    & SPAT1 Mutation Prediction & 1023 & Breast & TCGA \\
    & CDH1 Mutation Prediction & 1023 & Breast & TCGA \\
    \hline
     \multirow{2}*{LUAD/LUSC} & Subtype & 1053 & Lung & TCGA\\
    & LUSC TP53 Mutation Prediction & 408 & Lung & TCGA\\
    \hline
     \multirow{5}*{COAD/READ} & Overall Survival & 1318 & Colon & TCGA\\
    & TP53 Mutation Prediction & 1318 & Colon &TCGA \\
    & KRAS Mutation Prediction & 1318 & Colon &TCGA \\
    & Subtype & 1903 & Colorectal & TCGA \\
    \hline
    \end{tabular}
    \label{tab:datasets}
\end{table}
To train the foundation model, data from the TCGA, CPTAC, and PLCO cohorts were collected. All data is openly accessible, which enables reproducing our results and model. It also enables research towards biases within our model based on the input images.

TCGA and CPTAC data were obtained from the IDC. The IDC is a cloud-based repository of public cancer data, hosting images and analysis results from different modalities, including brightfield and fluorescence slide microscopy. All data in the IDC are harmonized into the Digital Imaging and Communications in Medicine (DICOM) format, versioned, and searchable by DICOM and non-DICOM metadata to facilitate transparency and reproducibility in imaging research studies \cite{fedorov2023national,schacherer2023nci,bontempi2024end}. We share the search query to obtain the TCGA and CPTAC data used in this study in the appendix queries \ref{cptac}, \ref{tcga}. Training data were obtained mainly from the TCGA data cohort, while the CPTAC cohort was held out for testing.

The PLCO cohort consists of 881 cases of radical prostatectomy (184 with event) with 2999 H\&E-stained WSIs. The cohort was cleaned to contain only cases with usable endpoints, leaving 875 (182) cases with 2352 slides. The cohort was then split (80/20) into a train and validation subset. Unfortunately, the PLCO cohort is not as easily accessible as the TCGA or CPTAC cohorts but requires an approved request.

From the TCGA and PLCO cohorts, we created 18 different tasks, listed in Table \ref{tab:datasets}, all of which had one label per slide. Subtyping targets were created through the TCGA-internal information, such as the diagnosed BRCA subtypes or diagnosed ISUP scores for prostate cancer. Survival and mutation targets were extracted from cBioPortal \cite{cerami2012cbio} for the corresponding patient IDs. We ensured that training and validation splits were coherent across tasks. For example, TCGA-NSCLC comprises the TCGA-LUAD and TCGA-LUSC cohorts, which share slides with the task for TCGA-LUSC-TP53 prediction. For both tasks, we ensured that the validation slides of one task were not included in the training set of the other task.

\subsection{Model training}
\begin{figure}
    \centering
    \includegraphics[width=\linewidth]{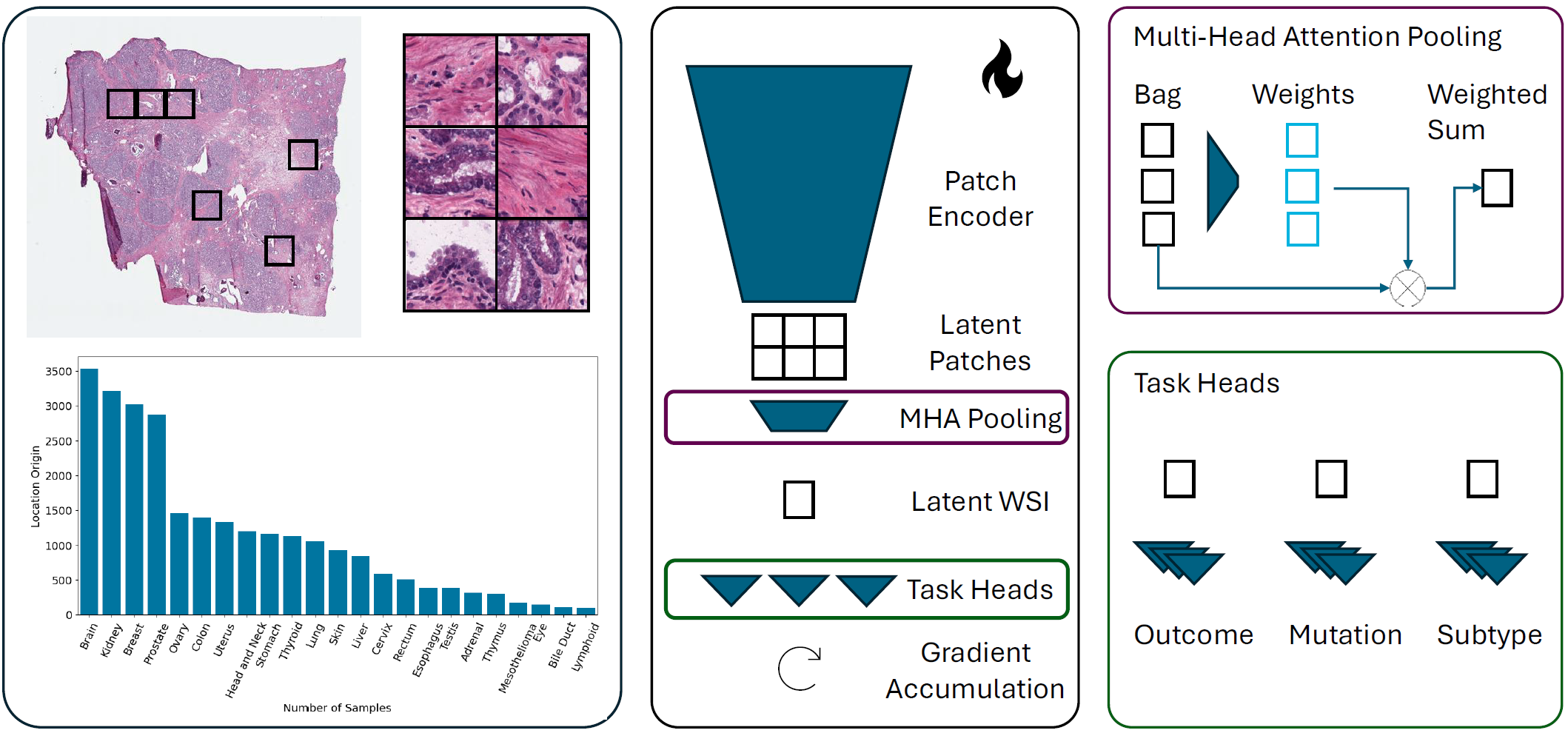}
    \caption{Schematic overview of the pipeline. A collection of WSI with corresponding labels (left) is used to train a tile encoder and pooling operation using multi-task learning (middle). The training is done in an end-to-end manner, iterating over individual tasks (lower right) and accumulating the gradient. Learnable weights rate each instance of the bag and then compress it into a latent WSI vector (top right).}
    \label{fig:figure_1}
\end{figure}
We propose a supervised, multi-task training scheme that is based on whole-slide-level labels from the TCGA, CPTAC, and PLCO datasets. Figure \ref{fig:figure_1} provides a schematic overview of the training strategy. Within-bag sampling allows for efficient end-to-end training of the tile encoder and the slide-level multi-head attention (MHA) pooling \cite{ilse2018attention}. Latent patches are rated with respect to their significance and then pooled by a weighted sum into a single latent WSI representation (Figure \ref{fig:figure_1} top right). Since the pooling operation is shared among all individual tasks, a general WSI representation can be learned such that all tasks can be solved individually from it. 

Each task was solved by a separate, small linear head. The head consisted of a 10\% dropout layer followed by a decision layer. Each training bag consisted of 128 to 256 randomly selected patches from a single WSI. All instances of one bag were passed through a shared encoder and pooled by the shared multi-head attention module, consisting of 8 heads.
This WSI vector was then used as input to a classification head based on the respective task. By iterating through the tasks and accumulating the individual losses, a combined optimizer step could be performed, solving all tasks simultaneously, as described in \cite{schafer2024overcoming}. As a backbone, a tiny Swin-Transformer V2 \cite{liu2022swin} was used and initialized with Tissue Concepts weights \cite{nicke2025tissue}.

During training, each patch in each bag was individually augmented using standard augmentation such as color shift, blurring, gray scaling, and deforming. Validation bags were not augmented and always contained 256 patches, the maximum size of a training bag.
The pipeline was trained for 200 epochs while monitoring the validation loss, and the best validation model was selected for testing. The entire pipeline was trained on one NVIDIA A100 for approximately 500 hours.

\subsection{Evaluation}
\label{subsec:eval}
\begin{table}[]
    \centering
    \caption{Overview of validation tasks.}
    \begin{tabular}{c|c|c|c}
        Task & Target & Organ & Domain \\
        \toprule
        Lung Cancer & Luad VS Lusc & Lung & In-domain \\
        Prostate Cancer & ISUP & Prostate & In-domain \\
        Breast Cancer & Benign vs Atypical vs Malignant & Breast & Out-of-domain \\
        Collagen Estimation & Low vs high fibrotic & Lung & Out-of-domain \\
        \midrule
        Prostate Survival &  Overall Survival & Prostate & In-domain \\
        LGG Survival & Overall Survival & Brain & Out-of-domain \\
        LUAD Survival & Overall Survival & Lung & Out-of-domain \\
        \botrule
    \end{tabular}
    \label{tab:eval_tasks}
\end{table}
For evaluation, we selected smaller cohorts based on challenges or openly available datasets. We created separate in- and out-of-domain evaluation tasks to fully assess the performance of the WSC model. All tasks are listed in Table \ref{tab:eval_tasks} with corresponding organ, target, and domain. 

\subsection{Evaluation protocol} \label{sec:eval_prot}
During evaluations, a linear decision layer was trained on the frozen encoder and  aggregator of the corresponding models.. For the patch-encoder model, we fine-tuned an ABMIL aggregator in combination with the decision layer. The fine-tuning was done over 100 epcohs with an adamW optimizer and a learning rate of 1e-4 while monitoring the validation loss. The model checkpoint with the lowest validation loss was used for testing. Each configuration was repeated 4 times to account for randomization artifacts during layer initialization and sample drawing.

\subsection{Evaluation datasets}
The CPTAC-NSCLC subset is used as an in-domain evaluation task (with TCGA-NSCLC used during pretraining).  It contains slides from 432 patients and was divided into train, validation, and test (40/10/50) splits. The task was to distinguish between lung adenocarcinoma and lung squamous cell carcinoma. Balanced accuracy and AUC were used as evaluation metrics.

The Prostate cANcer graDe Assessment (PANDA) challenge provides 10616 openly available WSIs from two different centers, Karolinska and Radboud \cite{bulten2022artificial}. A consensus ISUP score is assigned to each WSI. The score ranges from 0 (normal tissue) to 5 (severe prostate cancer). The evaluation task was to predict the ISUP score from the images. The dataset was split into training, validation, and testing (50/10/40), with a focus on balanced ISUP representations in each set. AUC and balanced accuracy were used as evaluation metrics.

The BReAst Cancer Subtyping (BRACS) dataset consists of 547 WSIs from 189 patients, which are already divided into training (395), validation (87), and test (65) splits \cite{brancati2022bracs}. The WSIs can be divided into three target classes: Benign, Atypical, and Malignant. The balanced accuracy and area under the ROC curve (AUC) were used as test metrics. This task contains classes that were not included in the pretraining and, therefore, is counted as an out-of-domain task.

The lung fibrosis estimation task is based on an internal dataset obtained at the Fraunhofer ITEM in Hannover. It includes 94 H\&E-stained precision-cut lung slices (PCLS) and corresponding tri-chrome-stained slides (split into 45/20/29 cases for training, validation, and testing) obtained and scanned at 20x from 10 donors. The tri-chrome-stained images were segmented using the semi-automatic software HistoKat \cite{hofener2019automated, arlt2016one}. The area percentage of collagen, lung parenchyma, other tissue, and background was computed on the segmentation masks. The same overall collagen distribution was assumed for the corresponding H\&E-stained slide pair, generating slide-level labels for each H\&E-stained slide. The dataset's median collagen area proportion was defined as a threshold for low or high collagen classes. We measured the AUC and balanced accuracy as evaluation metrics. Since this dataset assumes labels that were not included in the pretraining, it is used as an out-of-domain task.

The TCGA low-grade glioma cohort was used as one of three overall survival cohorts. The dataset consists of 480 cases, which were split into train, validation, and test sets (70/15/15).  It was ensured that enough unsensored cases are present within each split. Progress was measured in years, and each patient was grouped into one of four bins as a target label. The c-index was measured as the target metric.

The CPTAC LUAD cohort was used as one of three overall survival cohorts. The dataset consists of 157 cases, which were split into train, validation, and test sets (45/15/40). It was ensured that enough unsensored cases were present within each split. Progress was measured in years, and each patient was grouped into one of four bins as a target label. The c-index was measured as the target metric.

The training dataset of the Leopard Challenge \footnote{\url{https://leopard.grand-challenge.org/leopard/}} consists of 508 cases from Radboudumc with corresponding event time and censoring information. The dataset was split into train, validation, and test (50/10/40), where it was ensured that enough unsensored cases were present within each split. The overall time was divided into 4 distinct classes, which were used as target labels. The c-index was measured as the target metric.

\subsection{Ablation Study}
To find the best method for aggregation, we compared two common approaches. One was standard ABMIL with 8 heads, which was compared to an approach  where the top 5 patches voted for by each attention head were additionally used as input for the classification head. The Validation performance was used as a metric for comparison between the two approaches. 
We found that for patch-based classification and WSI-based classification tasks, the standard ABMIL head outperformed the ABMIL + top k approach.
We measured the performance on four of the training tasks with greater emphasis on weakly labeled tasks. Table \ref{tab:ablation_1} shows the maximum validation performance on three WSI-based tasks and one patch-based classification task. Looking at the patch-based classification task (CRC-WSI), the standard ABMIL head performed better over the ABMIL + Top k approach. The same pattern appears for WSI-based tasks. For the TCGA-NSCLC subtyping task, we find the AUC of 0.980 to be superior to 0.953. Similar performances can be seen for Brain tumor subtyping (LGG vs GBM) with 0.994 (ABMIL) and 0.968 (ABMIL + top k), and Kidney cancer subtyping (KICH vs KIRP vs KIRC).
\begin{table}[htbp]
    \caption{Comparison of different aggregation strategies with reported maximum AUC on validation set.}
    \centering
    \begin{tabular}{c|c|c}
        Task & ABMIL & ABMIL + Top k \\
        \toprule
        CRC-WSI & \textbf{0.994} & 0.981 \\
        KICH vs KIRP vs KIRC & \textbf{0.994} & 0.982 \\
        LGG vs GBM  & \textbf{0.994} & 0.968 \\
        NSCLC & \textbf{0.980} & 0.953 \\
        \botrule
    \end{tabular}
    \label{tab:ablation_1}
\end{table}

\backmatter


\bmhead{Acknowledgements}

This work was partially funded by the Fraunhofer-Gesellschaft through the Project FibroPaths.

The results published here are in part based on data generated by the TCGA Research Network: \url{https://www.cancer.gov/tcga}.

Data used in this publication were generated in part by the National Cancer Institute Clinical Proteomic Tumor Analysis Consortium (CPTAC).

The authors thank the National Cancer Institute for access to NCI's data collected by the Prostate, Lung, Colorectal, and Ovarian (PLCO) Cancer Screening Trial (CDAS PROJECT NUMBER: PLCOI-1612).










\bibliography{sn-bibliography}

@article{provgigapath,
  title={A whole-slide foundation model for digital pathology from real-world data},
  author={Xu, Hanwen and Usuyama, Naoto and Bagga, Jaspreet and Zhang, Sheng and Rao, Rajesh and Naumann, Tristan and Wong, Cliff and Gero, Zelalem and Gonz{\'a}lez, Javier and Gu, Yu and others},
  journal={Nature},
  volume={630},
  number={8015},
  pages={181--188},
  year={2024},
  publisher={Nature Publishing Group UK London}
}

@article{nicke2025tissue,
  title={Tissue concepts: Supervised foundation models in computational pathology},
  author={Nicke, Till and Sch{\"a}fer, Jan Raphael and H{\"o}fener, Henning and Feuerhake, Friedrich and Merhof, Dorit and Kie{\ss}ling, Fabian and Lotz, Johannes},
  journal={Computers in Biology and Medicine},
  volume={186},
  pages={109621},
  year={2025},
  publisher={Elsevier}
}

@article{schafer2024overcoming,
  title={Overcoming data scarcity in biomedical imaging with a foundational multi-task model},
  author={Sch{\"a}fer, Raphael and Nicke, Till and H{\"o}fener, Henning and Lange, Annkristin and Merhof, Dorit and Feuerhake, Friedrich and Schulz, Volkmar and Lotz, Johannes and Kiessling, Fabian},
  journal={Nature Computational Science},
  volume={4},
  number={7},
  pages={495--509},
  year={2024},
  publisher={Nature Publishing Group US New York}
}

@article{ding2024multimodal,
  title={Multimodal whole slide foundation model for pathology},
  author={Ding, Tong and Wagner, Sophia J and Song, Andrew H and Chen, Richard J and Lu, Ming Y and Zhang, Andrew and Vaidya, Anurag J and Jaume, Guillaume and Shaban, Muhammad and Kim, Ahrong and others},
  journal={arXiv preprint arXiv:2411.19666},
  year={2024}
}

@article{lenz2024unsupervised,
  title={Unsupervised Foundation Model-Agnostic Slide-Level Representation Learning},
  author={Lenz, Tim and Neidlinger, Peter and Ligero, Marta and W{\"o}lflein, Georg and van Treeck, Marko and Kather, Jakob Nikolas},
  journal={arXiv preprint arXiv:2411.13623},
  year={2024}
}

@article{shaikovski2024prism,
  title={Prism: A multi-modal generative foundation model for slide-level histopathology},
  author={Shaikovski, George and Casson, Adam and Severson, Kristen and Zimmermann, Eric and Wang, Yi Kan and Kunz, Jeremy D and Retamero, Juan A and Oakley, Gerard and Klimstra, David and Kanan, Christopher and others},
  journal={arXiv preprint arXiv:2405.10254},
  year={2024}
}

@phdthesis{hofener2019automated,
  title={Automated Quantification of Cellular Structures in Histological Images},
  author={H{\"o}fener, Henning},
  year={2019},
  school={Universit{\"a}t Bremen}
}

@article{brancati2022bracs,
  title={Bracs: A dataset for breast carcinoma subtyping in h\&e histology images},
  author={Brancati, Nadia and Anniciello, Anna Maria and Pati, Pushpak and Riccio, Daniel and Scognamiglio, Giosu{\`e} and Jaume, Guillaume and De Pietro, Giuseppe and Di Bonito, Maurizio and Foncubierta, Antonio and Botti, Gerardo and others},
  journal={Database},
  volume={2022},
  pages={baac093},
  year={2022},
  publisher={Oxford University Press UK}
}

@article{bulten2022artificial,
  title={Artificial intelligence for diagnosis and Gleason grading of prostate cancer: the PANDA challenge},
  author={Bulten, Wouter and Kartasalo, Kimmo and Chen, Po-Hsuan Cameron and Str{\"o}m, Peter and Pinckaers, Hans and Nagpal, Kunal and Cai, Yuannan and Steiner, David F and Van Boven, Hester and Vink, Robert and others},
  journal={Nature medicine},
  volume={28},
  number={1},
  pages={154--163},
  year={2022},
  publisher={Nature Publishing Group US New York}
}

@article{wang2024pathology,
  title={A pathology foundation model for cancer diagnosis and prognosis prediction},
  author={Wang, Xiyue and Zhao, Junhan and Marostica, Eliana and Yuan, Wei and Jin, Jietian and Zhang, Jiayu and Li, Ruijiang and Tang, Hongping and Wang, Kanran and Li, Yu and others},
  journal={Nature},
  volume={634},
  number={8035},
  pages={970--978},
  year={2024},
  publisher={Nature Publishing Group UK London}
}

@article{jiang2023mhattnsurv,
  title={MHAttnSurv: Multi-head attention for survival prediction using whole-slide pathology images},
  author={Jiang, Shuai and Suriawinata, Arief A and Hassanpour, Saeed},
  journal={Computers in biology and medicine},
  volume={158},
  pages={106883},
  year={2023},
  publisher={Elsevier}
}

@article{ilse2018attention,
  title={Attention-based deep multiple instance learning},
  author={Ilse, Maximilian and Tomczak, Jakub and Welling, Max},
  booktitle={International conference on machine learning},
  pages={2127--2136},
  year={2018},
  organization={PMLR},
}

@article{liu2022swin,
  title={Swin transformer v2: Scaling up capacity and resolution},
  author={Liu, Ze and Hu, Han and Lin, Yutong and Yao, Zhuliang and Xie, Zhenda and Wei, Yixuan and Ning, Jia and Cao, Yue and Zhang, Zheng and Dong, Li and others},
  booktitle={Proceedings of the IEEE/CVF conference on computer vision and pattern recognition},
  pages={12009--12019},
  year={2022}
}

@article{wang2022transformer,
  title={Transformer-based unsupervised contrastive learning for histopathological image classification},
  author={Wang, Xiyue and Yang, Sen and Zhang, Jun and Wang, Minghui and Zhang, Jing and Yang, Wei and Huang, Junzhou and Han, Xiao},
  journal={Medical image analysis},
  volume={81},
  pages={102559},
  year={2022},
  publisher={Elsevier}
}

@article{fedorov2021nci,
  title={NCI imaging data commons},
  author={Fedorov, Andrey and Longabaugh, William JR and Pot, David and Clunie, David A and Pieper, Steve and Aerts, Hugo JWL and Homeyer, Andr{\'e} and Lewis, Rob and Akbarzadeh, Afshin and Bontempi, Dennis and others},
  journal={Cancer research},
  volume={81},
  number={16},
  pages={4188--4193},
  year={2021},
  publisher={American Association for Cancer Research}
}

@article{fedorov2023national,
  title={National Cancer Institute Imaging Data Commons: toward transparency, reproducibility, and scalability in imaging artificial intelligence},
  author={Fedorov, Andrey and Longabaugh, William JR and Pot, David and Clunie, David A and Pieper, Steven D and Gibbs, David L and Bridge, Christopher and Herrmann, Markus D and Homeyer, Andr{\'e} and Lewis, Rob and others},
  journal={Radiographics},
  volume={43},
  number={12},
  pages={e230180},
  year={2023},
  publisher={Radiological Society of North America}
}

@article{schacherer2023nci,
  title={The NCI Imaging Data Commons as a platform for reproducible research in computational pathology},
  author={Schacherer, Daniela P and Herrmann, Markus D and Clunie, David A and H{\"o}fener, Henning and Clifford, William and Longabaugh, William JR and Pieper, Steve and Kikinis, Ron and Fedorov, Andrey and Homeyer, Andr{\'e}},
  journal={Computer methods and programs in biomedicine},
  volume={242},
  pages={107839},
  year={2023},
  publisher={Elsevier}
}

@article{bontempi2024end,
  title={End-to-end reproducible AI pipelines in radiology using the cloud},
  author={Bontempi, Dennis and Nuernberg, Leonard and Pai, Suraj and Krishnaswamy, Deepa and Thiriveedhi, Vamsi and Hosny, Ahmed and Mak, Raymond H and Farahani, Keyvan and Kikinis, Ron and Fedorov, Andrey and others},
  journal={Nature Communications},
  volume={15},
  number={1},
  pages={6931},
  year={2024},
  publisher={Nature Publishing Group UK London}
}

@article{cerami2012cbio,
  title={The cBio cancer genomics portal: an open platform for exploring multidimensional cancer genomics data},
  author={Cerami, Ethan and Gao, Jianjiong and Dogrusoz, Ugur and Gross, Benjamin E and Sumer, Selcuk Onur and Aksoy, B{\"u}lent Arman and Jacobsen, Anders and Byrne, Caitlin J and Heuer, Michael L and Larsson, Erik and others},
  journal={Cancer discovery},
  volume={2},
  number={5},
  pages={401--404},
  year={2012},
  publisher={American Association for Cancer Research}
}

@article{campanella2025clinical,
  title={A clinical benchmark of public self-supervised pathology foundation models},
  author={Campanella, Gabriele and Chen, Shengjia and Singh, Manbir and Verma, Ruchika and Muehlstedt, Silke and Zeng, Jennifer and Stock, Aryeh and Croken, Matt and Veremis, Brandon and Elmas, Abdulkadir and others},
  journal={Nature Communications},
  volume={16},
  number={1},
  pages={3640},
  year={2025},
  publisher={Nature Publishing Group UK London}
}

@article{arlt2016one,
  title={One size fits all: evaluation of the transferability of a new “learning” histologic image analysis application},
  author={Arlt, Janine and Homeyer, Andr{\'e} and S{\"a}nger, Constanze and Dahmen, Uta and Dirsch, Olaf},
  journal={Applied Immunohistochemistry \& Molecular Morphology},
  volume={24},
  number={1},
  pages={1--10},
  year={2016},
  publisher={LWW}
}

@article{chen2024towards,
  title={Towards a general-purpose foundation model for computational pathology},
  author={Chen, Richard J and Ding, Tong and Lu, Ming Y and Williamson, Drew FK and Jaume, Guillaume and Song, Andrew H and Chen, Bowen and Zhang, Andrew and Shao, Daniel and Shaban, Muhammad and others},
  journal={Nature Medicine},
  volume={30},
  number={3},
  pages={850--862},
  year={2024},
  publisher={Nature Publishing Group US New York}
}

@article{tang2025revisiting,
  title={Revisiting End-to-End Learning with Slide-level Supervision in Computational Pathology},
  author={Tang, Wenhao and Qin, Rong and Fang, Heng and Zhou, Fengtao and Chen, Hao and Li, Xiang and Cheng, Ming-Ming},
  journal={arXiv preprint arXiv:2506.02408},
  year={2025}
}

@article{pinckaers2021detection,
  title={Detection of prostate cancer in whole-slide images through end-to-end training with image-level labels},
  author={Pinckaers, Hans and Bulten, Wouter and Van der Laak, Jeroen and Litjens, Geert},
  journal={IEEE Transactions on medical imaging},
  volume={40},
  number={7},
  pages={1817--1826},
  year={2021},
  publisher={IEEE}
}

@article{gui2024survey,
  title={A survey on self-supervised learning: Algorithms, applications, and future trends},
  author={Gui, Jie and Chen, Tuo and Zhang, Jing and Cao, Qiong and Sun, Zhenan and Luo, Hao and Tao, Dacheng},
  journal={IEEE Transactions on Pattern Analysis and Machine Intelligence},
  volume={46},
  number={12},
  pages={9052--9071},
  year={2024},
  publisher={IEEE}
}

@article{ding2025multimodal,
  title={A multimodal whole-slide foundation model for pathology},
  author={Ding, Tong and Wagner, Sophia J and Song, Andrew H and Chen, Richard J and Lu, Ming Y and Zhang, Andrew and Vaidya, Anurag J and Jaume, Guillaume and Shaban, Muhammad and Kim, Ahrong and others},
  journal={Nature Medicine},
  pages={1--13},
  year={2025},
  publisher={Nature Publishing Group US New York}
}

@article{kheiri2025investigation,
  title={Investigation on potential bias factors in histopathology datasets},
  author={Kheiri, Farnaz and Rahnamayan, Shahryar and Makrehchi, Masoud and Asilian Bidgoli, Azam},
  journal={Scientific Reports},
  volume={15},
  number={1},
  pages={11349},
  year={2025},
  publisher={Nature Publishing Group UK London}
}

@article{kumari2025continual,
  title={Continual learning in medical image analysis: A comprehensive review of recent advancements and future prospects},
  author={Kumari, Pratibha and Chauhan, Joohi and Bozorgpour, Afshin and Huang, Boqiang and Azad, Reza and Merhof, Dorit},
  journal={Medical Image Analysis},
  pages={103730},
  year={2025},
  publisher={Elsevier}
}

@article{veta2019predicting,
  title={Predicting breast tumor proliferation from whole-slide images: the TUPAC16 challenge},
  author={Veta, Mitko and Heng, Yujing J and Stathonikos, Nikolas and Bejnordi, Babak Ehteshami and Beca, Francisco and Wollmann, Thomas and Rohr, Karl and Shah, Manan A and Wang, Dayong and Rousson, Mikael and others},
  journal={Medical image analysis},
  volume={54},
  pages={111--121},
  year={2019},
  publisher={Elsevier}
}

@phdthesis{bayasi2025beyond,
  title={Beyond catastrophic forgetting: advancing continual learning for robust and fair medical image analysis},
  author={Bayasi, Nourhan},
  year={2025},
  school={University of British Columbia}
}

@article{shi2025foundation,
  title={Foundation models: Insights and implications for gastrointestinal cancer},
  author={Shi, Lei and Huang, Rui and Zhao, Li-Ling and Guo, An-Jie},
  journal={World Journal of Gastroenterology},
  volume={31},
  number={47},
  pages={112921},
  year={2025}
}

@article{kather_predicting_2019,
	title = {Predicting survival from colorectal cancer histology slides using deep learning: {A} retrospective multicenter study},
	volume = {16},
	issn = {1549-1676},
	shorttitle = {Predicting survival from colorectal cancer histology slides using deep learning},
	url = {https://journals.plos.org/plosmedicine/article?id=10.1371/journal.pmed.1002730},
	doi = {10.1371/journal.pmed.1002730},
	language = {en},
	number = {1},
	urldate = {2024-03-08},
	journal = {PLOS Medicine},
	author = {Kather, Jakob Nikolas and Krisam, Johannes and Charoentong, Pornpimol and Luedde, Tom and Herpel, Esther and Weis, Cleo-Aron and Gaiser, Timo and Marx, Alexander and Valous, Nektarios A. and Ferber, Dyke and Jansen, Lina and Reyes-Aldasoro, Constantino Carlos and Zörnig, Inka and Jäger, Dirk and Brenner, Hermann and Chang-Claude, Jenny and Hoffmeister, Michael and Halama, Niels},
	month = jan,
	year = {2019},
	note = {Publisher: Public Library of Science},
	pages = {e1002730},
}

@article{spanhol_dataset_2016,
	title = {A {Dataset} for {Breast} {Cancer} {Histopathological} {Image} {Classification}},
	volume = {63},
	issn = {0018-9294, 1558-2531},
	url = {http://ieeexplore.ieee.org/document/7312934/},
	doi = {10.1109/TBME.2015.2496264},
	language = {en},
	number = {7},
	urldate = {2022-08-29},
	journal = {IEEE Transactions on Biomedical Engineering},
	author = {Spanhol, Fabio A. and Oliveira, Luiz S. and Petitjean, Caroline and Heutte, Laurent},
	month = jul,
	year = {2016},
	pages = {1455--1462},
}

@article{schomig-markiefka_quality_2021,
	title = {Quality control stress test for deep learning-based diagnostic model in digital pathology},
	volume = {34},
	issn = {08933952},
	url = {https://linkinghub.elsevier.com/retrieve/pii/S0893395222003702},
	doi = {10.1038/s41379-021-00859-x},
	language = {en},
	number = {12},
	urldate = {2023-07-13},
	journal = {Modern Pathology},
	author = {Schömig-Markiefka, Birgid and Pryalukhin, Alexey and Hulla, Wolfgang and Bychkov, Andrey and Fukuoka, Junya and Madabhushi, Anant and Achter, Viktor and Nieroda, Lech and Büttner, Reinhard and Quaas, Alexander and Tolkach, Yuri},
	month = dec,
	year = {2021},
	pages = {2098--2108},
}

@article{arvaniti_automated_2018,
	title = {Automated {Gleason} grading of prostate cancer tissue microarrays via deep learning},
	volume = {8},
	copyright = {2018 The Author(s)},
	issn = {2045-2322},
	url = {https://www.nature.com/articles/s41598-018-30535-1},
	doi = {10.1038/s41598-018-30535-1},
	language = {en},
	number = {1},
	urldate = {2024-03-08},
	journal = {Scientific Reports},
	author = {Arvaniti, Eirini and Fricker, Kim S. and Moret, Michael and Rupp, Niels and Hermanns, Thomas and Fankhauser, Christian and Wey, Norbert and Wild, Peter J. and Rüschoff, Jan H. and Claassen, Manfred},
	month = aug,
	year = {2018},
	note = {Publisher: Nature Publishing Group},
	pages = {12054},
}

@article{amgad_structured_2019,
	title = {Structured crowdsourcing enables convolutional segmentation of histology images},
	volume = {35},
	issn = {1367-4803},
	url = {https://doi.org/10.1093/bioinformatics/btz083},
	doi = {10.1093/bioinformatics/btz083},
	number = {18},
	urldate = {2024-03-08},
	journal = {Bioinformatics},
	author = {Amgad, Mohamed and Elfandy, Habiba and Hussein, Hagar and Atteya, Lamees A and Elsebaie, Mai A T and Abo Elnasr, Lamia S and Sakr, Rokia A and Salem, Hazem S E and Ismail, Ahmed F and Saad, Anas M and Ahmed, Joumana and Elsebaie, Maha A T and Rahman, Mustafijur and Ruhban, Inas A and Elgazar, Nada M and Alagha, Yahya and Osman, Mohamed H and Alhusseiny, Ahmed M and Khalaf, Mariam M and Younes, Abo-Alela F and Abdulkarim, Ali and Younes, Duaa M and Gadallah, Ahmed M and Elkashash, Ahmad M and Fala, Salma Y and Zaki, Basma M and Beezley, Jonathan and Chittajallu, Deepak R and Manthey, David and Gutman, David A and Cooper, Lee A D},
	month = sep,
	year = {2019},
	pages = {3461--3467},
}

@article{graham_mild-net_2019,
	title = {{MILD}-{Net}: {Minimal} {Information} {Loss} {Dilated} {Network} for {Gland} {Instance} {Segmentation} in {Colon} {Histology} {Images}},
	volume = {52},
	issn = {13618415},
	shorttitle = {{MILD}-{Net}},
	url = {http://arxiv.org/abs/1806.01963},
	doi = {10.1016/j.media.2018.12.001},
	language = {en},
	urldate = {2022-09-19},
	journal = {Medical Image Analysis},
	author = {Graham, Simon and Chen, Hao and Gamper, Jevgenij and Dou, Qi and Heng, Pheng-Ann and Snead, David and Tsang, Yee Wah and Rajpoot, Nasir},
	month = feb,
	year = {2019},
	note = {arXiv:1806.01963 [cs]},
	pages = {199--211},
}

@article{bulten_epithelium_2019,
	title = {Epithelium segmentation using deep learning in {H}\&{E}-stained prostate specimens with immunohistochemistry as reference standard},
	volume = {9},
	copyright = {2019 The Author(s)},
	issn = {2045-2322},
	url = {https://www.nature.com/articles/s41598-018-37257-4},
	doi = {10.1038/s41598-018-37257-4},
	language = {en},
	number = {1},
	urldate = {2024-03-08},
	journal = {Scientific Reports},
	author = {Bulten, Wouter and Bándi, Péter and Hoven, Jeffrey and Loo, Rob van de and Lotz, Johannes and Weiss, Nick and Laak, Jeroen van der and Ginneken, Bram van and Hulsbergen-van de Kaa, Christina and Litjens, Geert},
	month = jan,
	year = {2019},
	note = {Publisher: Nature Publishing Group},
	pages = {864},
}

@misc{shephard_tiager_2022,
	title = {{TIAger}: {Tumor}-{Infiltrating} {Lymphocyte} {Scoring} in {Breast} {Cancer} for the {TiGER} {Challenge}},
	shorttitle = {{TIAger}},
	url = {http://arxiv.org/abs/2206.11943},
	doi = {10.48550/arXiv.2206.11943},
	urldate = {2024-03-08},
	publisher = {arXiv},
	author = {Shephard, Adam and Jahanifar, Mostafa and Wang, Ruoyu and Dawood, Muhammad and Graham, Simon and Sidlauskas, Kastytis and Khurram, Syed Ali and Rajpoot, Nasir and Raza, Shan E. Ahmed},
	month = jun,
	year = {2022},
	note = {arXiv:2206.11943 [cs, eess]},
}

@article{crc-msi-dataset,
  title={Predicting microsatellite instability and key biomarkers in colorectal cancer from H\&E-stained images: achieving state-of-the-art predictive performance with fewer data using Swin Transformer},
  author={Guo, Bangwei and Li, Xingyu and Yang, Miaomiao and Jonnagaddala, Jitendra and Zhang, Hong and Xu, Xu Steven},
  journal={The Journal of Pathology: Clinical Research},
  volume={9},
  number={3},
  pages={223--235},
  year={2023},
  publisher={Wiley Online Library}
}

@misc{barbano2021unitopatho,
  author={Barbano, Carlo Alberto and Perlo, Daniele and Tartaglione, Enzo and Fiandrotti, Attilio and Bertero, Luca and Cassoni, Paola and Grangetto, Marco},
  booktitle={2021 IEEE International Conference on Image Processing (ICIP)}, 
  title={Unitopatho, A Labeled Histopathological Dataset for Colorectal Polyps Classification and Adenoma Dysplasia Grading}, 
  year={2021},
  volume={1},
  number={1},
  pages={76-80},
  doi={10.1109/ICIP42928.2021.9506198}
}

@article{saltz2018tumor,
  title={Tumor-infiltrating lymphocytes maps from tcga h\&e whole slide pathology images [data set]},
  author={Saltz, J and Gupta, R and Hou, L and Kurc, T and Singh, P and Nguyen, V and Samaras, D and Shroyer, KR and Zhao, T and Batiste, R and others},
  journal={Cancer Imaging Arch},
  volume={4},
  year={2018}
}

@article{abbet2022self,
  title={Self-rule to multi-adapt: Generalized multi-source feature learning using unsupervised domain adaptation for colorectal cancer tissue detection},
  author={Abbet, Christian and Studer, Linda and Fischer, Andreas and Dawson, Heather and Zlobec, Inti and Bozorgtabar, Behzad and Thiran, Jean-Philippe},
  journal={Medical image analysis},
  volume={79},
  pages={102473},
  year={2022},
  publisher={Elsevier}
}

@article{koziarski2024diagset,
  title={DiagSet: a dataset for prostate cancer histopathological image classification},
  author={Koziarski, Micha{\l} and Cyganek, Bogus{\l}aw and Niedziela, Przemys{\l}aw and Olborski, Bogus{\l}aw and Antosz, Zbigniew and {\.Z}ydak, Marcin and Kwolek, Bogdan and W{\k{a}}sowicz, Pawe{\l} and Buka{\l}a, Andrzej and Swad{\'z}ba, Jakub and others},
  journal={Scientific Reports},
  volume={14},
  number={1},
  pages={6780},
  year={2024},
  publisher={Nature Publishing Group UK London}
}

@article{nir2018automatic,
  title={Automatic grading of prostate cancer in digitized histopathology images: Learning from multiple experts},
  author={Nir, Guy and Hor, Soheil and Karimi, Davood and Fazli, Ladan and Skinnider, Brian F and Tavassoli, Peyman and Turbin, Dmitry and Villamil, Carlos F and Wang, Gang and Wilson, R Storey and others},
  journal={Medical image analysis},
  volume={50},
  pages={167--180},
  year={2018},
  publisher={Elsevier}
}

@article{silva2020going,
title={Going deeper through the Gleason scoring scale: An automatic end-to-end system for histology prostate grading and cribriform pattern detection},
author={Silva-Rodr{\'\i}guez, Julio and Colomer, Adri{\'a}n and Sales, Mar{\'\i}a A and Molina, Rafael and Naranjo, Valery},
journal={Computer Methods and Programs in Biomedicine},
volume={195},
pages={105637},
year={2020},
publisher={Elsevier}
}

@article{petrick2021spie,
  title={SPIE-AAPM-NCI BreastPathQ challenge: an image analysis challenge for quantitative tumor cellularity assessment in breast cancer histology images following neoadjuvant treatment},
  author={Petrick, Nicholas and Akbar, Shazia and Cha, Kenny H and Nofech-Mozes, Sharon and Sahiner, Berkman and Gavrielides, Marios A and Kalpathy-Cramer, Jayashree and Drukker, Karen and Martel, Anne L and BreastPathQ Challenge Group, for the},
  journal={Journal of Medical Imaging},
  volume={8},
  number={3},
  pages={034501--034501},
  year={2021},
  publisher={Society of Photo-Optical Instrumentation Engineers}
}

@article{weir2021cancer,
  title={Cancer incidence projections in the United States between 2015 and 2050},
  author={Weir, Hannah K and Thompson, Trevor D and Stewart, Sherri L and White, Mary C},
  journal={Preventing chronic disease},
  volume={18},
  pages={E59},
  year={2021}
}

@article{sung2021global,
  title={Global cancer statistics 2020: GLOBOCAN estimates of incidence and mortality worldwide for 36 cancers in 185 countries},
  author={Sung, Hyuna and Ferlay, Jacques and Siegel, Rebecca L and Laversanne, Mathieu and Soerjomataram, Isabelle and Jemal, Ahmedin and Bray, Freddie},
  journal={CA: a cancer journal for clinicians},
  volume={71},
  number={3},
  pages={209--249},
  year={2021},
  publisher={Wiley Online Library}
}

\begin{appendices}
\section{}\label{secA1}

\subsection{IDC cohorts}
We selected our cohorts of IDC data programmatically using the Python package idc-index\cite{fedorov2023national} version 0.8.2, which is installable via pip. To build the cohorts yourself, you can use the following code snippet and insert the respective queries: 

\begin{lstlisting}[language=Python]
from idc_index import index
idc_client = index.IDCClient()
idc_client.fetch_index("sm_instance_index")
idc_client.fetch_index("sm_index")
query = """<enter QUERY>"""   
our_cohort = idc_client.sql_query(query)
\end{lstlisting}

\begin{lstlisting}[language=SQL, caption=CPTAC-Query, label=cptac]
SELECT 
    collection_id as collection_id,
    index.PatientID as patient_id,
    index.StudyInstanceUID as study_id,
    sm_instance_index.SOPInstanceUID as slide_id,  
    sm_instance_index.SeriesInstanceUID as series_id,
    sm_instance_index.PixelSpacing_0 as pixel_spacing,
    CONCAT(
        TRIM(index.series_aws_url, '*'), 
        sm_instance_index.crdc_instance_uuid, 
        '.dcm'
    ) as url, 
    sm_index.primaryAnatomicStructureModifier_CodeMeaning as tissue
FROM 
    (sm_instance_index
    LEFT JOIN index 
        ON sm_instance_index.SeriesInstanceUID = index.SeriesInstanceUID)
    LEFT JOIN sm_index 
        ON sm_instance_index.SeriesInstanceUID = sm_index.SeriesInstanceUID 
WHERE
    index.Modality = 'SM'
    AND index.collection_id LIKE 'cptac_%'
    AND sm_instance_index.ImageType[3] = 'VOLUME'
    AND sm_index.illuminationType_code_designator_value_str = 'DCM:111744'
\end{lstlisting}

\begin{lstlisting}[language=SQL, caption=TCGA-Query, label=tcga]
SELECT 
    collection_id as collection_id,
    index.PatientID as patient_id,
    index.StudyInstanceUID as study_id,
    sm_instance_index.SOPInstanceUID as slide_id,  
    sm_instance_index.SeriesInstanceUID as series_id,
    sm_instance_index.PixelSpacing_0 as pixel_spacing,
    CONCAT(
        TRIM(index.series_aws_url, '*'), 
        sm_instance_index.crdc_instance_uuid, 
        '.dcm'
    ) as url, 
    sm_index.primaryAnatomicStructureModifier_CodeMeaning as tissue
FROM 
    (sm_instance_index
    LEFT JOIN index 
        ON sm_instance_index.SeriesInstanceUID = index.SeriesInstanceUID)
    LEFT JOIN sm_index 
        ON sm_instance_index.SeriesInstanceUID = sm_index.SeriesInstanceUID 
WHERE
    index.Modality = 'SM'
    AND index.collection_id LIKE 'tcga_%'
    AND sm_instance_index.ImageType[3] = 'VOLUME'
    AND sm_index.illuminationType_code_designator_value_str = 'DCM:111744'
\end{lstlisting}

\subsection{TC vs WSC}
We compared the starting point, Tissue Concepts, to the second version, WSC-tiny$_{wsi}$, on three of the evaluation tasks. Namely, Fibrosis estimation, BRACS, and NSCLC. The results can be seen in Table \ref{tab:ablation_2}. Our findings suggest that WSC-tiny$_{wsi}$ outperforms its predecessor on all tasks.
\begin{table}[htbp]
    \centering
    \caption{Comparison of TC and WSC-tiny$_{wsi}$ on three subtyping tasks. The average AUC is measured over 4 distinct runs.}
    \begin{tabular}{c|c|c}
    \toprule
        Task & TC & WSC-tiny$_{wsi}$ \\
        \midrule
        Fibrosis & 0.84 $\pm$ 0.02 &  \textbf{0.84} $\pm$ 0.02 \\
        BRACS & 0.70 $\pm$ 0.07 & \textbf{0.85} $\pm$ 0.01 \\
        NSCLC & 0.87 $\pm$ 0.06 & \textbf{0.96} $\pm$ 0.01\\
        \botrule
    \end{tabular}
    \label{tab:ablation_2}
\end{table}

\subsection{Overview of Patch-based tasks}
\begin{table}[h!]
    \caption{Overview of different patch-based tasks used in pre-training.}
    \centering
    \begin{tabular}{c|c|c}
    Cohort & Organ & Task Type\\
    \toprule
    NCT-CRC-HE 100K \cite{kather_predicting_2019} & Colorectal & Classification \\ 
    BreakHis \cite{spanhol_dataset_2016} & Breast & Classification \\
    CRC-MSI \cite{crc-msi-dataset} & Colorectal & Classification\\
    Diagset \cite{koziarski2024diagset} & Prostate & Classification \\
    Unitopatho \cite{barbano2021unitopatho} & Various & Classification \\
    Schoemig-Markiefka \cite{schomig-markiefka_quality_2021} & Prostate & Classification \\
    TCGA-TIL \cite{saltz2018tumor} & Various & Classification \\
    \midrule
    Arvaniti \cite{arvaniti_automated_2018} & Prostate & Segmentation \\
    BCSS \cite{amgad_structured_2019} & Breast & Segmentation \\
    Crag \cite{graham_mild-net_2019} & Colorectal & Segmentation \\
    CRC Phenotype \cite{abbet2022self} & Colorectal & Segmentation \\
    Gleason 19 \cite{nir2018automatic} & Prostate & Segmentation \\
    Peso \cite{bulten_epithelium_2019} & Prostate & Segmentation \\
    Sicap V2 \cite{silva2020going} & Prostate & Segmentation \\
    TIGER \cite{shephard_tiager_2022} & Breast & Segmentation \\
    \midrule
    BreastPathQ \cite{petrick2021spie} & Breast & Regression \\
    \botrule
    \end{tabular}
    \label{tab:patch-datasets}
\end{table}

\subsection{Result Tables}

\begin{table}[htbp]
    \centering
    \caption{Model comparison on in-domain tasks NSCLC-subtyping and ISUP prediction. The average area under the receiver operating characteristic curve (AUC) and balanced accuracy are reported across four distinct runs.}
   \begin{tabular}{c|c|c|c}
    \toprule
      Test Task & Model & AUC & b ACC\\
      \midrule
          \multirow{6}*{NSCLC} 
         & CHIEF & 0.97 $\pm$ 0.01 & 0.91 $\pm$ 0.01  \\ %
         & Prov-GigaPath & 0.97 $\pm$ 0.01 & 0.90 $\pm$ 0.02  \\ %
         & UNI & 0.97 $\pm$ 0.01 & 0.91 $\pm$ 0.01 \\
         & WSC-tiny$_{wsi}$ & 0.96 $\pm$ 0.01 & 0.905 $\pm$ 0.05 \\
         & WSC-tiny$_{wsi+patch}$ & \textbf{0.98} $\pm$ 0.01 & \textbf{0.92} $\pm$ 0.01 \\
         & WSC-small$_{wsi+patch}$ & 0.97 $\pm$ 0.01 & 0.91 $\pm$ 0.01\\
         \midrule
         \multirow{6}*{ISUP prediction} 
         & CHIEF & 0.83 $\pm$ 0.01 & 0.46 $\pm$ 0.02\\ %
         & Prov-GigaPath & \textbf{0.89} $\pm$ 0.01 & \textbf{0.57} $\pm$ 0.01  \\
         & UNI & 0.89 $\pm$ 0.01 & 0.57 $\pm$ 0.01  \\
         & WSC-tiny$_{wsi}$ & 0.88 $\pm$ 0.02 & 0.51 $\pm$ 0.01 \\
         & WSC-tiny$_{wsi+patch}$ & \textbf{0.89} $\pm$ 0.01 & \textbf{0.57} $\pm$ 0.02 \\
         & WSC-small$_{wsi+patch}$ & 0.88 $\pm$ 0.01 & 0.54 $\pm$ 0.01 \\
        \botrule
    \end{tabular}
    \label{tab:inDomain}
\end{table}

\begin{table}[htbp]
    \centering
    \caption{Model comparison on out-of-domain tasks BRCAS and Fibrosis estimation. Average Area under the receiver operator curve (AUC) and balanced accuracy over four distinct runs are reported.}
   \begin{tabular}{c|c|c|c}
    \toprule
      Test Task & Model & AUC & b ACC\\
      \midrule
          \multirow{6}*{BRACS} 
         & CHIEF & 0.83 $\pm$ 0.01 & 0.58 $\pm$ 0.02 \\ 
         & Prov-GigaPath & 0.71 $\pm$ 0.04 & 0.41 $\pm$ 0.06 \\
         & UNI & 0.82 $\pm$ 0.01 & 0.61 $\pm$ 0.03\\
         & WSC-tiny$_{wsi}$ & \textbf{0.85} $\pm$ 0.01 & \textbf{0.63} $\pm$ 0.02\\
         & WSC-tiny$_{wsi+patch}$ & 0.81 $\pm$ 0.01 & 0.60 $\pm$ 0.04\\
         & WSC-small$_{wsi+patch}$ & 0.84 $\pm$ 0.02 & 0.63 $\pm$ 0.05 \\
         \midrule
         \multirow{6}*{Fibrosis}
         & CHIEF & 0.83 $\pm$ 0.01 & 0.58 $\pm$ 0.06 \\ 
         & Prov-GigaPath & 0.75 $\pm$ 0.09 & 0.69 $\pm$ 0.1 \\
         & UNI & 0.72 $\pm$ 0.09 & 0.64 $\pm$ 0.07\\
         & WSC-tiny$_{wsi}$ & \textbf{0.84} $\pm$ 0.02 & 0.67 $\pm$ 0.07\\
         & WSC-tiny$_{wsi+patch}$ & 0.80 $\pm$ 0.06 & 0.70 $\pm$ 0.06\\
         & WSC-small$_{wsi+patch}$ & 0.82 $\pm$ 0.06 & \textbf{0.71} $\pm$ 0.06 \\
        \botrule
    \end{tabular}
    \label{tab:outOfDomain}
\end{table}

\begin{table}[htbp]
 \caption{Model comparison on outcome prediction tasks of Low-Grade Glioma, Clear Cell Renal Cell Carcinoma, and Lung Adenocarcinoma. The average c-index over four distinct runs and the standard deviation are reported.}
    \centering
   \begin{tabular}{c|c|c}
    \toprule
      Test Task & Model & C-index \\
      \midrule
          \multirow{6}*{TCGA-LGG} 
         & CHIEF & 0.61 $\pm$ 0.01 \\
         & Prov-GigaPath  &  0.65 $\pm$ 0.03 \\
         & UNI  &  0.51 $\pm$ 0.05 \\
         & WSC-tiny$_{wsi}$  &  0.61 $\pm$ 0.07 \\
         & WSC-tiny$_{wsi+patch}$ & 0.64 $\pm$ 0.04 \\
         & WSC-small$_{wsi+patch}$ & \textbf{0.66} $\pm$ 0.02\\
         \midrule
         \multirow{6}*{CPTAC-LUAD OS} 
         & CHIEF & \textbf{0.70} $\pm$ 0.01 \\
         & Prov-GigaPath  &  0.62 $\pm$ 0.02 \\
         & UNI  &  0.63 $\pm$ 0.02 \\
         & WSC-tiny$_{wsi}$  &  0.55 $\pm$ 0.02 \\
         & WSC-tiny$_{wsi+patch}$ & 0.64 $\pm$ 0.02 \\
         & WSC-small$_{wsi+patch}$ & 0.60 $\pm$ 0.02 \\
         \midrule
         \multirow{6}*{LEOPARD} 
         & CHIEF & 0.58 $\pm$ 0.01 \\
         & Prov-GigaPath  & 0.54 $\pm$ 0.01 \\
         & UNI  & 0.55 $\pm$ 0.01 \\
         & WSC-tiny$_{wsi}$  & 0.56 $\pm$ 0.03 \\
         & WSC-tiny$_{wsi+patch}$ & \textbf{0.66} $\pm$ 0.03 \\
         & WSC-small$_{wsi+patch}$ & 0.60 $\pm$ 0.07 \\
         \botrule
    \end{tabular}
    \label{tab:survival}
\end{table}

\subsection{Heatmaps}


\begin{figure}
    \centering
    \includegraphics[width=0.99\linewidth]{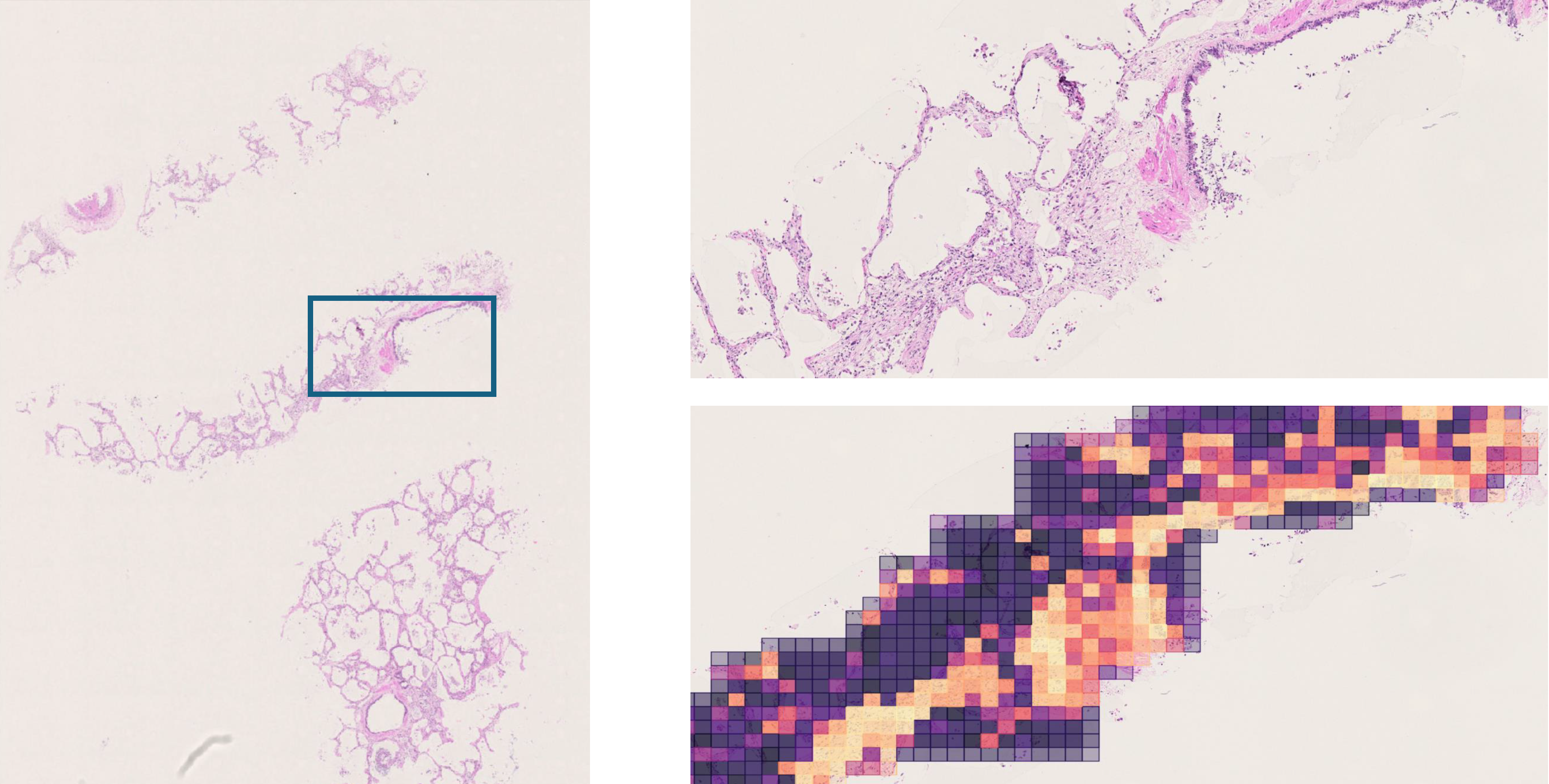}
    \caption{Attention weight visualization of a model fine-tuned for lung fibrosis estimation.}
    \label{fig:attention_weights}
\end{figure}




\end{appendices}


\end{document}